\documentclass[doublecol]{epl2} 
\usepackage{amsmath}
\usepackage{amssymb}
\usepackage{amsthm}
\usepackage{mathtools}
\usepackage{braket}
\usepackage[colorlinks=true]{hyperref}
 
\DeclareMathOperator{\Tr}{Tr}

\title{Non-Gaussianity of random quantum states}
\shorttitle{Non-Gaussianity of random quantum states} 

\author{Filiberto Ares\inst{1} \and Sara Murciano\inst{2} \and Pasquale Calabrese\inst{1}}
\shortauthor{F. Ares, S. Murciano, P. Calabrese}

\institute{                    
  \inst{1} SISSA and INFN, via Bonomea 265, 34136 Trieste, Italy\\
  \inst{2} Universit\'e Paris-Saclay, CNRS, LPTMS, 91405, Orsay, France
}

\abstract{We study the fermionic non-Gaussianity in typical quantum states, focusing on Haar random states of qubits with or without a global $U(1)$ symmetry. Using the Weingarten calculus, we derive analytical predictions for the non-Gaussianity, defined as the relative entropy between the reduced density matrix and its Gaussianized counterpart. We identify two regimes controlled by the ratio between the subsystem and the system size, $\ell/L$. For $\ell/L < 1/2$, the non-Gaussianity vanishes in the absence of symmetries, because typical reduced density matrices are exponentially close to the maximally mixed state. In the presence of a global $U(1)$ symmetry, instead, it remains small but finite. 
By contrast, in the regime $\ell/L > 1/2$, the non-Gaussianity becomes extensive. These results establish the typical scaling of fermionic non-Gaussianity in random states and analyze how this is modified by the presence of global symmetries. }

\begin{document}

\maketitle

\section{Introduction} 
Quantum resources provide a unifying framework for characterizing 
and quantifying properties of quantum systems~\cite{cg-19}. In this approach, one
defines a restricted set of “free” operations and corresponding “free”
states that can be prepared without cost, while states outside this set
are considered resources. Their resource content is quantified by
monotones that measure their deviation from the set of free states. A
paradigmatic example is entanglement, where separable states are the free
states, and local operations with classical communication constitute the allowed
operations; entanglement entropy is the canonical monotone for pure
states, capturing correlations and phases of matter~\cite{hhhh-09, afov-08,ccd-09,l-16}. 
Beyond entanglement, other resources have recently gained attention in 
extended quantum systems, as they provide complementary insights into the 
rich structure of complex many-body states. These include 
magic~\cite{vhge-14, lw-22, loh-22}, which 
quantifies the non-stabilizerness essential for universal quantum computation, 
and asymmetry~\cite{vaw-08, brs-07, gms-09, ms-14, amc-23}, which measures symmetry breaking and has proven crucial in 
revealing non-equilibrium phenomena such as the quantum Mpemba effect~\cite{acm-25}.

Another relevant quantum resource is fermionic non-Gaussianity, for which 
the free states are Gaussian~\cite{bravyi-05}, i.e. they satisfy Wick's theorem. These states 
play a central role in 
many-body physics, as they naturally arise in non-interacting 
systems~\cite{gaudin-60, lsm-61}. 
Despite their apparent simplicity, they exhibit many nontrivial properties 
that also appear in more complex systems and often capture the essential 
physical features of interest. Their structure enables the use of
powerful analytical techniques and efficient numerical methods that are 
typically inaccessible in interacting models. As a result, fermionic 
Gaussian states underlie many approximate methods in condensed 
matter physics~\cite{st-22}. In quantum computation, matchgate circuits can be mapped 
onto free fermions, allowing efficient classical simulation~\cite{tdv-02}. 
In analogy to Clifford circuits, where the inclusion of non-Clifford 
gates enables universal quantum computation, matchgate circuits can
achieve universality by incorporating suitable non-Gaussian operations~\cite{brod-16, hjksy-19}.

This context motivates investigating the extent to which a given state 
deviates from Gaussianity. In this work, we analyze fermionic 
non-Gaussianity in ensembles of random states of qubit systems. 
Random quantum states capture typical properties of quantum systems, 
making them a powerful tool for studying fermionic non-Gaussianity 
in generic settings. In particular, random quantum states are expected to describe the qualitative behavior of the excited eigenstates of typical Hamiltonians and to capture the long-time universal features of sufficiently chaotic dynamics~\cite{vr-17, vhbr-17, bhkrv-22}. They are also relevant to the problem of black hole evaporation, where the entropy of the emitted radiation would qualitatively follow the Page curve, that is, the curve described by the average entanglement entropy of Haar-random states, provided that information is conserved~\cite{aemm-19, penington-20}.
Here, we consider two kinds of ensembles: Haar-random 
states, which are uniformly sampled over the full Hilbert space and lack 
any symmetry, and Haar-random states with an additional 
global $U(1)$ symmetry. Entanglement entropy~\cite{page-93-1, page-93-2, bianchi19, bdm-26}, symmetry resolved entanglement~\cite{mcp-22, lntt-22, bdk-24}, negativity~\cite{bn-13, slk-21,cvv-24}, asymmetry~\cite{ampc-24, rac-25, rac-25-2}, and non-stabilizerness~\cite{tds-25, ievh-25, tts-25, irjoh-26} have been studied in these particular ensembles, including in recent experiments on a quantum simulator~\cite{yjahzc-26}.
To quantify non-Gaussianity, we compute the 
average relative entropy with respect to their Gaussianized counterparts, 
a widely used resource monotone~\cite{gpb-08, gp-10, mm-13, lumia-24, lb-24, csg-25, asst-25, ats-25, ammscp-26}.
We remark that alternative measures of non-Gaussianity exist in the literature~\cite{gm-05, tmpp-17, pp-18, dk-24, cs-23, rs-24, bmel-25, sierant26, nzs-26}, although they will not be discussed in the present work. 

\section{Fermionic non-Gaussianity} 

Consider a system of $L$ qubits in a state described by a density matrix 
$\rho$. Via the Jordan-Wigner transformation, the Pauli operators at each 
qubit $j$ define the fermionic operators $c_j$, $c_j^\dagger$ obeying the 
canonical anticommutation relations $\{c_j^\dagger, c_{j'}\}=\delta_{jj'}$ 
and $\{c_j, c_{j'}\}=0$. 
The fermionic two-point correlation functions 
$C_{nm} = {\rm Tr}(\rho\, c_n^\dagger c_m)$ and 
$F_{nm} = {\rm Tr}(\rho\, c_n c_m)$ uniquely determine the Gaussianization of $\rho$, which can be written as~\cite{peschel-03}
\begin{equation}\label{eq:gaussianized_rho}
\rho_G=\frac{1}{Z}e^{-\mathbf{c}^\dagger K \mathbf{c}}, \quad K=\log\frac{I-\Gamma}{I+\Gamma},
\end{equation}
where $Z=\mathrm{Tr}(e^{-\mathbf{c}^\dagger K \mathbf{c}})$, $\mathbf{c}=(c_1, \dots, c_L, c_1^\dagger, \dots, c_L^\dagger)$
and $\Gamma$ is the $2L\times 2L$ matrix
\begin{equation}
\Gamma=\left(\begin{array}{cc} 2C-I & 2F^\dagger \\ 2F & I-2C^T\end{array}\right).
\end{equation} 
The non-Gaussianity of $\rho$ is then defined as the relative entropy 
between $\rho$ and $\rho_G$
\begin{equation}
{\rm NG}(\rho)=S(\rho || \rho_G)={\rm Tr}(\rho(\log\rho-\log\rho_G)).
\end{equation}
By definition, ${\rm NG}(\rho)\geq 0$, and it vanishes if and only if $\rho=\rho_G$, i.e., when $\rho$ is Gaussian.
Since ${\rm Tr}(\rho\log\rho_G)={\rm Tr}(\rho_G\log\rho_G)$, the non-Gaussianity can be rewritten
as the difference of von Neumann entropies
 \begin{equation}\label{eq:NG_diff}
{\rm NG}(\rho)=S(\rho_G)-S(\rho),
\end{equation}
where $S(\rho)=-{\rm Tr}(\rho \log \rho)$. In particular, using the specific form~\eqref{eq:gaussianized_rho} of $\rho_G$, its
entropy can be computed from the correlation matrix $\Gamma$ through the well-known formula~\cite{peschel-03, vidal-03}
\begin{equation}\label{eq:gauss_rho_A}
S(\rho_G) = -\frac{1}{2}\mathrm{Tr}\left[\frac{I+\Gamma}{2}\log\frac{I+\Gamma}{2} 
+ \frac{I-\Gamma}{2}\log\frac{I-\Gamma}{2}\right]
\end{equation}

\section{Haar random states}
Let us start by analyzing the non-Gaussianity of Haar-random states. 
We take as local basis of each qubit $j$ the 
states $\{\ket{0}_j, \ket{1}_j\}$, with $j=1,\dots, L$. The ensemble of Haar-random
states is constructed by uniformly sampling over the total Hilbert space $\mathcal{H}$; 
thus, it consists of pure states of the form
$U\ket{0}$ where $U$ is a $D\times D$, with $D=2^L$, random unitary matrix and 
$\ket{0}=\otimes_j \ket{0}_j$. We want to study the non-Gaussianity of the 
reduced density matrix $\rho_A={\Tr}_B(U\ket{0}\bra{0}U^\dagger)$ in a subsystem $A$ of $\ell$ qubits. From the definition of non-Gaussianity~\eqref{eq:NG_diff}, it is clear that to obtain its average value,
we need to compute the average entanglement entropy of the state $\rho_A$ as well as  of their Gaussianization $\rho_{A, G}$.
The former is very well known and gives the  Page curve~\cite{page-93-1, page-93-2}. We report here its final expression,
 \begin{equation}\label{eq:page_curve}
 \mathbb{E}[S(\rho_A)]=\Psi(2^L + 1) - 
 \Psi(2^{L - \ell} + 1) - \frac{2^\ell - 1}{2^{L - \ell + 1}}, \quad \ell<L/2,
\end{equation}  
where $\Psi(z)$ is the Digamma function. For $\ell > L/2$, the corresponding expression follows from the substitution $\ell \mapsto L-\ell$.
This result was first conjectured by Page \cite{page-93-1, page-93-2} and later rigorously proved in different ways \cite{fk-94, sen-96}.

To our knowledge, the entanglement entropy of the reduced 
Gaussianized Haar-random states $\rho_{A,G}$ has not been calculated. To 
obtain it, we can expand Eq.~\eqref{eq:gauss_rho_A} in 
the moments of the two-point  correlation matrix $\Gamma$ restricted to 
$A$,
\begin{equation}\label{eq:exp_f}
 \mathbb{E}[S(\rho_{A, {\rm G}})]= \ell \log 2 - \frac{1}{2}\sum_{k=1}^\infty \frac{\mathbb{E}[{\rm Tr}(\Gamma_A^{2k})]}{2k(2k-1)}.
 \end{equation}
In Ref.~\cite{sierant26}, a general formula for the leading-order behavior of $\mathbb{E}[{\rm Tr}(\Gamma_A^{2k})]$ was derived in the limit $\ell = L$ and $L \to \infty$. 
In the present case, however, we also require the subleading contributions for generic values of $\ell$. 
In the appropriate limits, our results reduce to those of Ref.~\cite{sierant26}.

To compute  the average value of the moments of $\Gamma_A$, we are going to 
use standard tools of random matrix theory. To this end, let us write the random states as 
\begin{equation}
U\ket{0}=\sum_{a=0}^{D-1} U_a \ket{a},
\end{equation}
where $U_a$ are the entries of the first column of the matrix $U$, i.e. $U_a=U_{a, 0}$, and 
$\ket{a}$ denote all the elements of the basis of $\mathcal{H}$ constructed from the local basis
$\{\ket{0}_j, \ket{1}_j\}$. It is also useful to introduce the Majorana fermions $\gamma_{2j-1}=c_j^\dagger+c_j$
and $\gamma_{2j}=i(c_j^\dagger-c_j)$. Their two-point correlation matrix $G_{j, j'}=i\langle \gamma_j \gamma_{j'}\rangle$
is related to $\Gamma_A$ via $\Gamma_A=-V(iG_A+I)V^\dagger$, where $V$ is a unitary transformation. Then we can write the average of the first two moments of $G_A$ as
\begin{equation}\label{eq:tr_GA}
\mathbb{E}[{\rm Tr}(G_A)]= i\sum_{j=1}^{2\ell} \sum_{a, a'=0}^{D-1}\mathbb{E}[U_a U_{a'}^*]\bra{a'} \gamma_j \gamma_{j}\ket{a},
\end{equation}
and
\begin{multline}\label{eq:tr_GA2}
\mathbb{E}[{\rm Tr}(G_A^2)]= -\sum_{j, j'=1}^{2\ell} \sum_{a, a', b, b'=0}^{D-1} \mathbb{E}[U_a U_b U_{a'}^*U_{b'}^*]
\\
\times \bra{a'} \gamma_j \gamma_{j'}\ket{a}\bra{b'} \gamma_{j'} \gamma_{j}\ket{b}.
\end{multline}
Analogous expressions can be straightforwardly written for the higher moments. The averages of the form 
$\mathbb{E}[U_{a_1}\cdots U_{a_k}U_{a_1'}^*\dots U_{a_k'}^*]$ can
be systematically obtain through the Weingarten formula~\cite{weingarten-78, cs-06}
\begin{equation}\label{eq:weingarten}
\mathbb{E}[U_{a_1}\dots U_{a_k} U^*_{a'_1}\dots U^*_{a'_k}]
= \sum_{\sigma,\tau\in S_k} \mathrm{Wg}_k(\sigma\tau^{-1}) \prod_{i=1}^k \delta_{a_i,a'_{\sigma(i)}},
\end{equation}
where $S_k$ is the symmetric group and $\mathrm{Wg}_k(\sigma)$ are the Weingarten coefficients. They
satisfy
\begin{equation}\label{eq:weingarten_sum}
\sum_{\tau\in S_k} \mathrm{Wg}_k(\tau)=\frac{(D-1)!}{(D+k-1)!}.
\end{equation}
\begin{figure}[t]
\centering
 \includegraphics[width=0.49\textwidth]{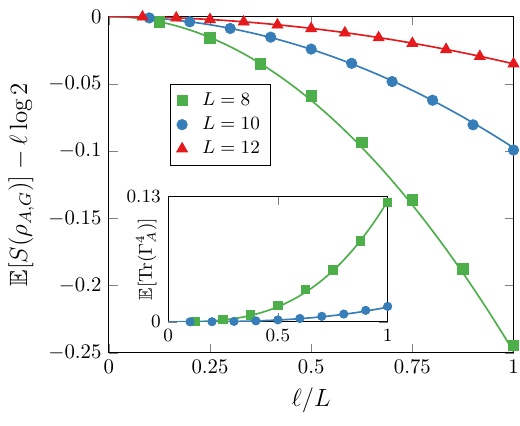}
 \caption{Average entanglement entropy of the Gaussianized reduced density matrix $\rho_{A, G}$ for Haar random states as a function 
 of the subsystem size $\ell$ in a system of $L$ qubits. The symbols are the exact average over 100 random states.
 The solid lines are the analytic prediction in Eq.~\eqref{eq:final_ent_rhoAG}. In the inset, we plot $\mathbb{E}[{\rm Tr}(\Gamma_A^4)]$, which determines the 
 $O(2^{-2L})$ correction to Eq.~\eqref{eq:final_ent_rhoAG}. Symbols are the exact average over the same random states as in the main plot. Solid curves are Eq.~\eqref{eq:av_Gamma4_Haar}.}
 \label{fig:s_rhoG_haar}
 \end{figure}
For the first two moments~\eqref{eq:tr_GA} and~\eqref{eq:tr_GA2}, it takes the exact particular form
\begin{align}\label{eq:wein}
\mathbb{E}[U_a U_{a'}^*]=\frac{\delta_{aa'}}{D},\quad
\mathbb{E}[U_a U_b U_{a'}^* U_{b'}^*]=\frac{\delta_{aa'}\delta_{bb'}+\delta_{ab'}\delta_{ba'}}{D(D+1)}.
\end{align}
Since Majorana fermions obey $\{\gamma_j, \gamma_{j'}\}=2\delta_{jj'}$, 
then Eqs.~\eqref{eq:tr_GA} and~\eqref{eq:tr_GA2} readily give
\begin{align}
\mathbb{E}[{\rm Tr}(G_A)]=i2\ell,\quad \mathbb{E}[{\rm Tr}(G_A^2)]=-2\ell-\frac{2\ell(2\ell-1)}{D+1},
\end{align}
and, therefore,
\begin{equation}\label{eq:ev_Gamma_2_haar}
\mathbb{E}[{\rm Tr}(\Gamma_A^2)]=\frac{2\ell(2\ell-1)}{D+1}.
\end{equation}
Truncating at this order in the expansion~\eqref{eq:exp_f}, we obtain
\begin{equation}\label{eq:final_ent_rhoAG}
\mathbb{E}[S(\rho_{A, G})]=\ell\log(2)-\frac{\ell(2\ell-1)}{2(2^L+1)}+O(2^{-2L}).
\end{equation}

We check numerically the accuracy of this expression in Fig.~\ref{fig:s_rhoG_haar}. The symbols are the 
exact average entanglement entropy of the Gaussianized reduced density matrix in an ensemble of
100 Haar random states. The curves correspond to the result in Eq.~\eqref{eq:final_ent_rhoAG}. We obtain an excellent agreement. The $O(2^{-2L})$ correction in Eq.~\eqref{eq:final_ent_rhoAG} is determined by $\mathbb{E}[{\rm Tr}(\Gamma_A^4)]$. 
Using Eqs.~\eqref{eq:weingarten}-\eqref{eq:weingarten_sum},  we find 
\begin{multline}\label{eq:av_Gamma4_Haar}
\mathbb{E}[{\rm Tr}(\Gamma_A^4)]
=
\frac{
(28+2D)\ell
-16(3+D)\ell^2
}{
(1+D)(2+D)(3+D)
}
\\
+
\frac{
16(D-2)\ell^3
+32\ell^4
}{
(1+D)(2+D)(3+D)
}
\stackrel{\ell,L\to\infty}\to \frac{16\ell^3}{2^{2L}}.
\end{multline}
In the inset of Fig.~\ref{fig:s_rhoG_haar}, we check this result using the same ensemble of Haar random states as in the main plot.

\begin{figure}[t]
\centering
 \includegraphics[width=0.49\textwidth]{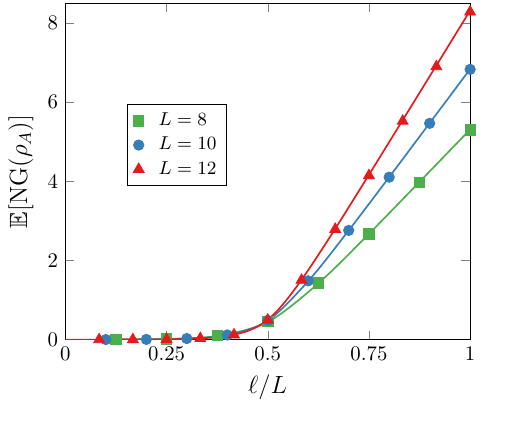}
 \caption{Average non-Gaussianity for Haar random states of a subsystem of size 
 $\ell$ in a system of $L$ qubits. The symbols are the exact average over 100 random states.
 The solid lines are the analytic prediction using Eqs.~\eqref{eq:page_curve} and~\eqref{eq:final_ent_rhoAG}.}
 \label{fig:av_ng_haar}
 \end{figure}

Finally, combining Eqs.~\eqref{eq:page_curve} and~\eqref{eq:final_ent_rhoAG}, we obtain the average non-Gaussianity of $\rho_A$ in the 
Haar random ensemble at leading order. In Fig.~\ref{fig:av_ng_haar}, we plot the result as a function of $\ell/L$ for different values 
of $L$ and compare with the exact average for an ensemble of 100 Haar random states. We obtain an 
excellent agreement. 

In the thermodynamic limit $L\to\infty$ with $\ell/L$ finite, the 
non-Gaussianity behaves as
\begin{equation}\label{eq:ng_haar_th_lim}
\mathbb{E}[{\rm NG}(\rho_A)]=\left\{\begin{array}{ll}
0, & \ell<L/2,\\
(2\ell-L)\log2, &  \ell>L/2.
\end{array}\right.
\end{equation}
Remarkably, this result implies that $\rho_A$ is effectively Gaussian in the thermodynamic limit whenever $\ell < L/2$. This follows from the decoupling inequality~\cite{ph-07}
\begin{equation}\label{eq:ph-dec}
\mathbb{E}\!\left[\left\|\rho_A-\frac{I}{2^{\ell}}\right\|_1\right]^2
\leq 2^{2\ell-L},
\end{equation}
where $\|\bullet\|_1$ denotes the trace norm. Therefore, when $\ell<L/2$, $\rho_A$ is exponentially close on average to the infinite-temperature state (i.e. the normalized identity), which is Gaussian. On the other hand, when $\ell>L/2$, the inequality~\eqref{eq:ph-dec} does not provide a useful bound because the rhs is exponentially large. In the thermodynamic limit, the average non-Gaussianity~\eqref{eq:ng_haar_th_lim} is a continuous function but becomes nonanalytic at the Page time, $\ell=L/2$, where its derivative with respect to the subsystem size exhibits a jump discontinuity. The origin of this singularity lies in the average entropy of $\rho_A$: in the thermodynamic limit, Eq.~\eqref{eq:page_curve} behaves as $\mathbb{E}[S(\rho_A)]\sim \ell\log 2$ for $\ell<L/2$ and as $\mathbb{E}[S(\rho_A)]\sim (L-\ell)\log 2$ for $\ell>L/2$, whereas the average entropy of its Gaussianization~\eqref{eq:final_ent_rhoAG} is smooth and behaves as $\mathbb{E}[S(\rho_{A,G})]\sim \ell\log 2$ for all $\ell$.

This behavior contrasts with that of the asymmetry. Indeed, the asymmetry of $\rho_A$ also vanishes for $\ell<L/2$ in the thermodynamic limit as a consequence of the decoupling inequality~\eqref{eq:ph-dec}; however, at the Page time $\ell=L/2$ it exhibits a discontinuous jump to a finite value~\cite{ampc-24, rac-25}.

Finally, a formula analogous to Eq.~\eqref{eq:gauss_rho_A} exists for the R\'enyi entanglement entropy of $\rho_{A, G}$, $S_n(\rho_{A, G})=\log{\rm Tr}(\rho_{A, G}^n)/(1-n)$,
\begin{equation}
S_n(\rho_{A, G})=\frac{1}{2(1-n)}{\rm Tr}
\log\left[\left(\frac{I+\Gamma_A}{2}\right)^n+\left(\frac{I-\Gamma_A}{2}\right)^n\right].
\end{equation}
Expanding the logarithm on the right hand side and truncating it at second order, 
\begin{equation}\label{eq:renyi_exp_haar}
\mathbb{E}[S_n(\rho_{A, G})]\approx
\ell \log 2 + \frac{a_2(n)}{1-n}  
\mathbb{E}[{\rm Tr}(\Gamma_A^2)]+\frac{a_4(n)}{1-n}\mathbb{E}[{\rm Tr}(\Gamma_A^4)],
\end{equation}
where $a_2(n)=n(n-1)/4$ and $a_4(n)=(-3n+4n^2-n^4)/24$. Applying Eq.~\eqref{eq:ev_Gamma_2_haar}, we obtain
\begin{equation}
\mathbb{E}[S_n(\rho_{A, G})]\approx
\ell \log 2 - \frac{n\ell(2\ell-1)}{2(2^L+1)}+O(2^{-2L}).
\end{equation}
Since the R\'enyi entropies are accessible in quantum simulators through randomized measurements \cite{efh-23}, replacing the von Neumann entropy in Eq.~\eqref{eq:NG_diff} with the R\'enyi entropies provides a probe of the non-Gaussianity of experimental many-body states.

\section{Spectrum of $\rho_{A, G}$}
\begin{figure}[t]
\centering
 \includegraphics[width=0.49\textwidth]{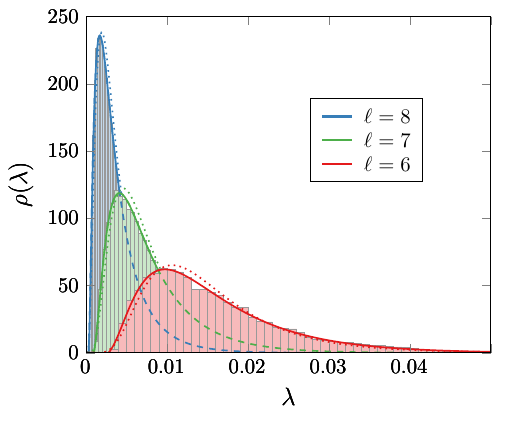}
 \caption{Eigenvalue distribution of the Gaussianized reduced density matrix $\rho_{A, G}$ of Haar-random states in a system of $L=8$ qubits and different subsystem sizes $\ell$. The histograms show the distribution of the exact eigenvalues obtained from $10^4$ random-state samples. The solid curves represent the approximation in Eqs.~\eqref{eq:app_moments_rhoAG}-\eqref{eq:spectrum}. Dotted curves correspond to the log-normal distribution~\eqref{eq:log-normal}.}
 \label{fig:spectrum}
 \end{figure}

The knowledge of the R\'enyi entropies of $\rho_{A, G}$ also allows us to characterize the distribution 
$\rho(\lambda)=2^{-\ell}\sum_j\delta(\lambda-\lambda_j)$ of its eigenvalues 
$\lambda_j$~\cite{cl-08}. 
Introducing the normalized moments $\mu_n=\mathbb{E}[{\rm Tr}(\rho_{A, G}^n)]/2^\ell$, then
\begin{equation}\label{eq:mu_n}
\mu_n=\int_0^\infty d\lambda\, \lambda^n \rho(\lambda).
\end{equation}
 Using Eq.~\eqref{eq:renyi_exp_haar} and assuming that $\mathbb{E}[e^a]\approx e^{\mathbb{E}[a]}$, we have 
\begin{equation}\label{eq:app_moments_rhoAG}
\mu_n\approx e^{-n\ell\log2+a_2(n)\mathbb{E}[{\rm Tr}(\Gamma_A^2)]+a_4(n)\mathbb{E}[{\rm Tr}(\Gamma_A^4)]}.
\end{equation}
Performing the change of variables $\lambda = e^x$ in Eq.~\eqref{eq:renyi_exp_haar}, doing the analytic continuation
$n=ik$, and applying the inverse Fourier transform, we obtain the inverse relation
\begin{equation}\label{eq:spectrum}
\rho(\lambda)=\frac{1}{2\pi \lambda}\int_{-\infty}^\infty 
dk\, e^{-ik\log\lambda}\mu_{ik}.
\end{equation}
If we neglect the $O(2^{-L})$ term, i.e. the term $\mathbb{E}[{\rm Tr}(\Gamma_A^4)]$, in Eq.~\eqref{eq:app_moments_rhoAG},
then $\rho(\lambda)$ is the log-normal distribution
\begin{equation}\label{eq:log-normal}
\rho(\lambda)\approx \frac{e^{-(\log\lambda + \ell\log2 + \sigma)^2/(4 \sigma)}}{2\lambda\sqrt{\pi \sigma}} ,\quad \sigma=\mathbb{E}[{\rm Tr}(\Gamma_A^2)]/4.
\end{equation}
In Fig.~\ref{fig:spectrum}, we verify these results for $L=8$ qubits. The histograms show the distribution of the exact eigenvalues of $\rho_{A,G}$ obtained by sampling $10^4$ random states. These eigenvalues were calculated numerically from the eigenvalues of $\Gamma_A$, as described in Ref.~\cite{cl-08}. The solid curves correspond to Eq.~\eqref{eq:spectrum} using the approximation in Eq.~\eqref{eq:app_moments_rhoAG}, while dotted ones correspond to the log-normal distribution~\eqref{eq:log-normal}. For the system size considered and $\ell > L/2$, good agreement with the numerical results is achieved only after including the term $\mathbb{E}[{\rm Tr}(\Gamma_A^4)]$ in Eq.~\eqref{eq:av_Gamma4_Haar}, as clear from Fig. \ref{fig:spectrum}. For $\ell < L/2$, additional subleading corrections to the approximation in Eq.~\eqref{eq:app_moments_rhoAG} must be taken into account to obtain a good description of the data for $L=8$ or slightly larger.

 \section{$U(1)$ Haar random states} 
 We now consider Haar random states with global $U(1)$ charge. We can decompose the Hilbert space as $\mathcal{H}=\bigoplus_{M=0}^L \mathcal{H}(M)$, where $\mathcal{H}(M)\subset\mathcal{H}$ is the charge eigenspace associated with the eigenvalue $M$. We denote its dimension by $D=\binom{L}{M}$ and we consider random states $\ket{\psi}$ drawn from the uniform Haar
measure on $\mathcal{H}(M)$. Since the total charge is fixed, the Gaussianized state is particle-number conserving, and it is therefore completely determined by the correlation matrix $C$. Its entanglement entropy can be obtained by setting $\Gamma=2C-I$ in Eq.~\eqref{eq:gauss_rho_A}. 
Let $\nu=M/L$ be the filling. The analogue of Eq.~\eqref{eq:exp_f} is obtained by
expanding around the average correlation matrix $\nu I_A$. Indeed, we find
\begin{equation}\label{eq:expansion}
 \begin{split}
\mathbb{E}[S(\rho_{A,\mathrm{G}})]
=
\ell H(\nu)
+
H'(\nu)\mathbb{E}[\Tr(C_A-\nu I_A)]\\+
\sum_{m=2}^{\infty}
\frac{H^{(m)}(\nu)}{m!}
\mathbb{E}[\Tr\left[(C_A-\nu I_A)^m] \right],
 \end{split}
 \end{equation}
where $H(x)=-x\log x-(1-x)\log(1-x)$. To get the expansion at leading order in the system size, we focus on the first and second moments. Using Eqs.~\eqref{eq:wein}, we notice that the linear term vanishes because
\begin{equation}
\mathbb E[\Tr C_A]=\ell \nu.
\end{equation}
Thus, the first non-trivial correction is given by the second moment
\begin{equation}
\begin{split}
\mathbb E[\mathrm{Tr}(C_A^2)]
=&
\frac{1}{D(D+1)}
\Big[
\ell \binom{L-1}{M-1}^2\\
+&
\ell \binom{L-1}{M-1}
+
\ell(\ell-1)\binom{L-2}{M-1}
\Big].
\end{split}
\end{equation} 
Plugging these expressions in Eq.~\eqref{eq:expansion}, we obtain 
\begin{equation}\label{eq:toplot}
\begin{split}
  \mathbb{E}[S(\rho_{A,\mathrm{G}})]
=&    \ell H(\nu)+\frac{1}{2}H''(\nu)\Big(\frac{1}{D(D+1)}
\Big[
\ell \binom{L-1}{M-1}^2\\+&\ell \binom{L-1}{M-1}
+
\ell(\ell-1)\binom{L-2}{M-1}
\Big]-\ell\nu^2\Big).
\end{split}
\end{equation}
\begin{figure}[t]
\centering
 \includegraphics[width=0.49\textwidth]{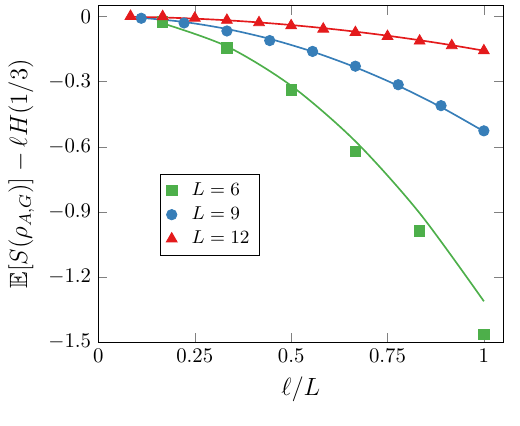}
 \caption{Average entanglement entropy of the Gaussianized reduced density matrix $\rho_{A, G}$ for $U(1)$ Haar random states as a function 
 of the subsystem size $\ell$ in a system of $L$ qubits at filling $\nu=1/3$. The symbols are the exact average over 100 random states.
 The solid lines are the analytic prediction in Eq.~\eqref{eq:toplot}.}
 \label{fig:s_rhoG_haaru1}
 \end{figure}
Hence, the Gaussianized entropy is extensive,
up to exponentially small corrections in the dimension of the 
Hilbert space. Fig.~\ref{fig:s_rhoG_haaru1} shows the deviation of the Gaussianized entropy from its extensive contribution, $\ell H(\nu)$, at fixed filling $\nu=1/3$.
The agreement with Eq.~\eqref{eq:toplot} confirms that the leading correction is already captured by the second moment of the
correlation matrix. 

\begin{figure}[t]
\centering
\includegraphics[width=0.49\textwidth]{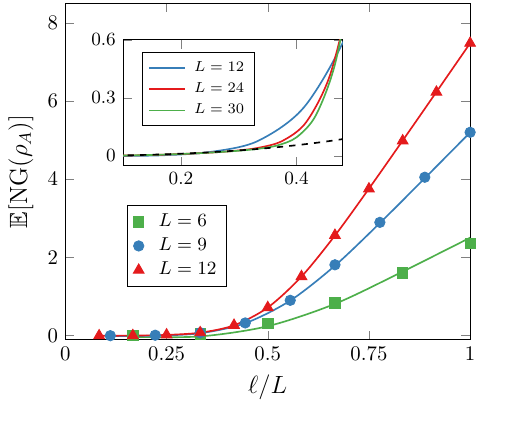}
\caption{Average non-Gaussianity for $U(1)$ Haar random states of a subsystem of size 
$\ell$ in a system of $L$ qubits at filling $\nu=1/3$. The symbols are the exact average over 100 random states.
The solid lines are the analytic prediction using Eqs.~\eqref{eq:toplot} and~\eqref{eq:toplot2}. The inset zooms into the regime $\ell < L/2$, illustrating how the agreement with Eq.~\eqref{eq:NGapp} (black dashed line) improves with increasing $L$. }
\label{fig:av_ng_haaru1}
\end{figure}

\begin{figure}[t]
\centering
\includegraphics[width=0.49\textwidth]{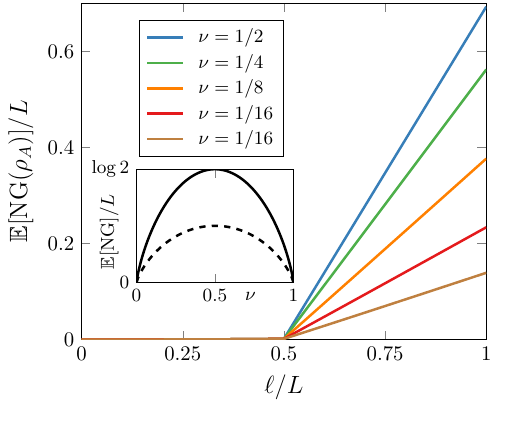}
\caption{Average non-Gaussianity~\eqref{eq:NGapp} for $U(1)$ symmetric Haar-random states at different fillings $\nu$ in the thermodynamic limit $L\to\infty$, with  $\ell/L$ finite. For $\ell<L/2$, the non-Gaussianity is not extensive in $L$ and, unlike in the absence of any symmetry, is a small but finite function of $\ell/L$ that is independent of $\nu$. The inset shows the dependence on $\nu$ for $\ell/L=3/4$ (dashed curve) and $\ell=L$ (solid curve).}
\label{fig:av_ng_haaru1_thermo}
\end{figure}

The average non-Gaussianity is then obtained by subtracting the average entropy of the reduced density matrix. In terms of 
\begin{equation}
    d_A(q)=\binom{\ell}{q},\quad 
d_B(M-q)=\binom{L-\ell}{M-q},
\end{equation}
and $D_q=d_A(q)d_B(M-q)$, the average entropy of the subsystem can be written as~\cite{bianchi19}
\begin{equation}\label{eq:toplot2}
\mathbb E[S(\rho_A)]
=\Psi(D+1)-\sum_q\frac{D_q}{D}
\left[\Psi(d_1(q)+1)
+\frac{d_2(q)-1}{2d_1(q)}
\right]
\end{equation}
where $d_1(q)=\max\{d_A(q),d_B(M-q)\}$ and $d_2(q)=\min\{d_A(q),d_B(M-q)\}$. Combining these expressions gives the leading order prediction for the average
non-Gaussianity of $U(1)$ symmetric Haar-random states at fixed $\nu$ and $\ell/L$,
\begin{equation}\label{eq:NGapp}
\begin{split}
&\mathbb{E}[{\rm NG}(\rho_A)]=\\
&\left\{\begin{array}{ll}
-1/2(\ell/L+\log(1-\ell/L)), & \ell<L/2,\\
(2\ell-L)H(\nu)-1/2(1-\ell/L+\log(\ell/L)), &  \ell>L/2.
\end{array}\right.
\end{split}
\end{equation}
Thus, in contrast to the Haar ensemble, the average
non-Gaussianity of $U(1)$ Haar-random states is subextensive for $\ell<L/2$ and approaches a small but finite function of the subsystem fraction. Fig.~\ref{fig:av_ng_haaru1} displays the average non-Gaussianity in the $U(1)$ symmetric Haar ensemble at filling $\nu=1/3$ and different system sizes. The agreement between the numerical data and the analytic prediction obtained from Eqs.~\eqref{eq:toplot} and \eqref{eq:toplot2} shows that the non-Gaussianity is characterized by two different regimes: for \(\ell<L/2\), the extensive terms cancel, and the non-Gaussianity approaches an $O(1)$ function of $\ell/L$ independent of the filling $\nu$ (see the inset). For $\ell>L/2$, instead, the Gaussianized entropy produces an extensive contribution to the non-Gaussianity, which depends on $\nu$, as illustrated in Fig.~\ref{fig:av_ng_haaru1_thermo}.

 \section{Conclusion}

In this work, we characterized the amount of non-Gaussianity in typical random states, both in the absence and in the presence of a global $U(1)$ symmetry. In both cases, we identified two qualitatively distinct regimes. For $\ell < L/2$, the non-Gaussianity vanishes exactly for typical states in the thermodynamic limit, since such states are locally indistinguishable from the maximally mixed state. In the presence of a global $U(1)$ symmetry, however, a residual structure survives in the reduced density matrix, leading to a small but finite non-Gaussianity. By contrast, for $\ell > L/2$, the non-Gaussianity becomes extensive and exhibits a volume-law scaling. Exactly at $\ell = L/2$, the non-Gaussianity remains continuous in the thermodynamic limit, while developing a nonanalytic behavior.
 
Our results open several directions for future investigation. 
For instance, in the context of pure fermionic states, a different measure of non-Gaussianity has recently been shown to detect quantum critical points~\cite{sierant26}. It would be very interesting to extend this perspective to mixed states, for example by analyzing the non-Gaussianity of reduced density matrices, along the lines developed in the present work.
Measurement-induced phenomena provide a particularly promising setting in which to address these questions~\cite{potter22,fisher23}. In the absence of measurements, the non-Gaussianity of typical states is extensive, similarly to several other quantities that have been employed to characterize measurement-induced transitions. It would therefore be interesting to understand how this volume-law behavior is modified, suppressed, or ultimately destroyed by the presence of measurements, potentially providing a new diagnostic of measurement-induced phenomena. 
Another natural direction concerns the role of non-Gaussianity in quantum state preparation. Since fermionic Gaussian states can be prepared efficiently \cite{mh-25}, the scaling of non-Gaussianity may provide useful information about the complexity required to prepare typical many-body states or interacting quantum states.

\section{Acknowledgments} The authors thank Xhek Turkeshi for useful discussions. P.C. and F.A. acknowledge support from the European Research Council under the Advanced Grant no. 101199196 (MOSE).

\end{document}